\documentclass[a4paper,usenatbib]{mnras}

\usepackage{amsmath}
\usepackage{amssymb}
\usepackage{graphicx}
\usepackage[normalem]{ulem}

\usepackage[switch]{lineno}
\usepackage[dvipsnames]{xcolor}

\title[Ground-based GRB detection prospetcs]{Gamma-Ray Burst detection prospects for next generation ground-based VHE facilities}

\author[G. La Mura et al.]{G. La Mura\thanks{\textit{Contacts}: glamura@lip.pt, swgo\_spokespersons@swgo.org},$^1$
 U. Barres de Almeida,$^2$
 R. Concei\c{c}\~ao,$^{1,3}$
 A. De Angelis,$^{4,5,6}$ \newauthor
 F. Longo,$^{7,8,9}$
 M. Pimenta,$^{1,3}$
 E. Prandini, $^{6,10}$
 E. Ruiz-Velasco,$^{11}$
 B. Tom\'e$^{1,3}$\\
 $^1$Laborat\'orio de Instrumenta\c{c}\~ao e F\'{i}sica Experimental de Part\'{i}culas (LIP), Av. Prof. Gama Pinto 2, 1649-003 Lisboa, Portugal\\
 $^2$Brazilian Center for Physics Research (CBPF), Rua Dr. Xavier Sigaud 150, 22290-180 Rio de Janeiro, Brazil\\
 $^3$Instituto Superior T\'ecnico (IST), Av. Rovisco Pais 1, 1049-001 Lisboa, Portugal\\
 $^4$Dipartimento di Fisica e Astronomia - Universit\`a di Padova, Via Marzolo 8, 35131 Padova, Italy\\
 $^5$Dipartimento di scienze matematiche, informatiche e fisiche - Universit\`a degli Studi di Udine, Via Palladio 8, 33100 Udine, Italy\\
 $^6$Istituto Nazionale di Fisica Nucleare sez. Padova (INFN), Via Marzolo 8, 35131 Padova, Italy\\
 $^7$IFPU - Institute for Fundamental Physics of the Universe, Via Beirut 2, 34014 Trieste, Italy\\
 $^8$INFN, Sezione di Trieste, via A. Valerio 2, 34100 Trieste, Italy\\
 $^9$Dipartimento di Fisica, Universit\`a degli Studi di Trieste, via A. Valerio 2, 34100 Trieste, Italy\\
 $^{10}$INAF - Osservatorio Astronomico di Padova, Vicolo dell'Osservatorio 3, 35122 Padova, Italy\\
 $^{11}$Max-Planck-Institut f\"ur Kernphysik, P.O. Box 103980, 69029 Heidelberg, Germany}

\date{Received July 28$^{th}$, 2021. Accepted September 6$^{th}$, 2021; in original form July 28$^{th}$, 2021.}
\pubyear{2021}

\begin{document}
\newcommand{\de}{\mathrm{d}}
\newcommand{\ergsec}{\mathrm{erg\, s^{-1}}}
\newcommand{\ergcmsec}{\mathrm{erg\, cm^{-2}\, s^{-1}}}
\newcommand{\ergcm}{\mathrm{erg\, cm^{-2}}}
\newcommand{\usec}{\mathrm{s}}
\maketitle
\begin{abstract}
    Gamma-ray Bursts (GRB) were discovered by satellite-based detectors as powerful sources of transient $\gamma$-ray emission. The Fermi satellite detected an increasing number of these events with its dedicated Gamma-ray Burst Monitor (GBM), some of which were associated with high energy photons ($E > 10\,$GeV), by the Large Area Telescope (LAT). More recently, follow-up observations by Cherenkov telescopes detected very high energy emission ($E > 100\,$GeV) from GRBs, opening up a new observational window with implications on the interpretation of their central engines and on the propagation of very energetic photons across the Universe. Here, we use the data published in the 2nd Fermi-LAT Gamma Ray Burst Catalogue to characterise the duration, luminosity, redshift and light curve of the high energy GRB emission. We extrapolate these properties to the very high energy domain, comparing the results with available observations and with the potential of future instruments. We use observed and simulated GRB populations to estimate the chances of detection with wide-ﬁeld ground-based $\gamma$-ray instruments. Our analysis aims to evaluate the opportunities of the Southern Wide-ﬁeld-of-view Gamma-ray Observatory (SWGO), to be installed in the Southern Hemisphere, to complement CTA. We show that a low-energy observing threshold ($E_{low} < 200\,$GeV), with good point source sensitivity ($F_{lim} \approx 10^{-11} \ergcmsec$ in $1\,$yr), are optimal requirements to work as a GRB trigger facility and to probe the burst spectral properties down to time scales as short as $10\,$s, accessing a time domain that will not be available to IACT instruments.
\end{abstract}


\section{Introduction}
Gamma-ray Bursts (GRB) are the most powerful electromagnetic events known in the Universe. Although they were discovered at the dawn of space exploration with satellite-based detectors \citep{Klebesadel73}, their recent observation at very high energies \citep[VHE, $E > 100\,$GeV,][]{GRBpaper, GRBpaper2}, or in coincidence with Gravitational Wave emission \citep[GW,][]{Abbott17a, Abbott17b}, raised new interest within the scientific community. Their isotropic distribution soon pointed to an extra-galactic origin \citep{Fishman95}, but reliable estimates on their intrinsic power were only possible when multi-wavelength observations permitted the firm identification of counterparts, leading to measurements of their redshift \citep[e.g.][]{Metzger97,Kulkarni98}. The observed radiation fluxes indicate intrinsic isotropic luminosities falling in the range between $10^{48}\, \ergsec \leq L_{iso} \leq 10^{52}\, \ergsec$. In spite of their remarkable beaming, such luminosities still lie orders of magnitude above the average power radiated by a whole galaxy. The most natural explanation for the emission of similar amounts of energy is to associate these events with the disruptive processes that end the life of very massive stars ($M \geq 20\, \mathrm{M_\odot}$) or follow the merger of compact stellar remnants, especially neutron stars, leading to the formation of a {\it magnetar} or a black hole \citep{Woosley93}. Such events are purportedly accompanied by the acceleration of an ultra-relativistic jet of plasma, where matter and radiation interact at the highest energy scales, producing a {\it prompt} flash of $\gamma$-ray photons, followed by a decaying phase named {\it afterglow}. The prompt stage is highly variable and it appears to be composed by pulse-like signals \citep{Nakar02}. The afterglow, on the contrary, is a smoothly evolving function of time.

It was very soon realized that the distribution of GRBs presented bimodal characteristics \citep{Mazets81, Norris84}. Depending on the duration of the prompt emission, GRBs are divided into two families: a short class, with a prompt phase lasting less than $2\,$s, and a long one, having more extended prompt emission \citep{Kouveliotou93}. The different characteristics of these families are in good agreement with two distinct types of source. Long GRBs are best explained by the core-collapse mechanisms of very massive stars and they can actually be observed in connection with supernova explosions \citep{Galama98, Bloom99, Cano16}. Short GRBs, on the contrary, nicely fit in the binary compact object merger scenario \citep{Eichler89, Li98, Abbott17a}, as eventually confirmed by the GRB~170817A - GW~170817 multi-messenger association \citep{Abbott17b, Blanchard17}. This distinction has considerable implications also on the properties of the afterglow. The long-lasting emission, indeed, is connected with the expansion and cooling of the blast and the production of energetic photons is favoured by its possible interactions with the surrounding environment \citep{Gompertz18}. It is generally expected that long GRBs occur in a denser environment, with respect to short ones, due to the intense mass loss processes that the stellar progenitor probably experienced ahead of its final stages of evolution \citep{Fruchter06}. In addition, massive stars evolve quickly and they generally end their existence close to an actively star forming region. Short GRBs, on the contrary, require longer time-scales to be triggered, since this involves the formation of a compact binary stellar remnant and its subsequent merger. However, evidence of High Energy emission (HE, $E \geq 10\,$GeV), collected by the {\it Fermi} Large Area Telescope \citep[{\it Fermi}-LAT,][]{LATpaper}, suggests that energetic radiation is not exclusively due to external interactions, since it can be detected in both event types and also in association with the prompt emission \citep[e.g.][]{Asano09}. VHE observations will be crucial to properly identify the emission mechanisms, due to the discriminating power that the energy and timing of these photons have on jet structure models, physics of relativistic shocks and potential hadronic contributions. It is very likely that future Imaging Atmospheric Cherenkov Telescopes (IACT) observations, particularly with the Cherenkov Telescope Array \citep[CTA,][]{CTApaper}, will further clarify the role of VHE emission in the afterglow of GRBs, but their investigation during the prompt stage requires a different approach. Indeed, observations of fast transients like GRBs are challenging for pointing instruments like IACTs, which can only operate after an accurate positional trigger has located the source. Due to the fast evolution of the prompt emission, this implies an unavoidable loss of information. This problem is particularly relevant since many questions, concerning the particle acceleration sites and the dominant radiating species, could be only answered through VHE observations of the bursts at early times.

In this paper we investigate the possibilities of new generation monitoring instruments, based on extensive air shower (EAS) detector arrays, such as the Southern Wide-field-of-view Gamma-ray Observatory \citep[SWGO,][]{SWGOpaper}, to probe the early stages of GRB emission at VHE and to trigger follow-up observations, thus complementing the role of IACTs. To this purpose, we analyse the performances of different realistic configurations, and we estimate the expected scientific impact. In our study, we use data collected over 10 years of observations in the $2^{\rm nd}$ {\it Fermi}-LAT GRB Catalogue \citep[2FLGC,][]{2FLGCpaper} to address the observational requirements that would most effectively probe the nature of VHE emission in GRBs. We use a combination of observed properties and synthetic distributions to evaluate the GRB detection probabilities and spectral reconstruction capabilities of an instrument like SWGO over a range of integration times and energies, and we evaluate how these complement the capabilities of current and planned IACT instruments. The paper is structured as follows: in \S2, we present the  data collection, together with the simulations used to derive the synthetic GRB properties adopted in the study; in \S3, we evaluate the potential that SWGO has to contribute in interpreting long and short GRBs, comparing it with that of other instruments; in \S4, we discuss the effects that different instrument performance sets have on the visibility of targets; finally, in \S5, we present our conclusions. We base our calculations on a flat Universe with $\Lambda$CDM Cosmology, adopting $H_0 = 69.6\, \mathrm{km\, s^{-1}\, Mpc^{-1}}$, $\Omega_\Lambda = 0.714$ and $\Omega_M = 0.286$ \citep{Bennett14, PlanckColl16}.

\section{High-energy data and simulations}
\subsection{Data selection}
One of the distinguishing features of GRBs is their peculiar temporal evolution. The prompt stage, where the bulk of their energy is emitted between $100\,$keV and $100\,$MeV, is characterized by high variability \citep{Nakar02}. Higher energy photons are generally observed with some delay or during the afterglow, though the detection of HE photons associated with the variable prompt phase has been reported in some cases \citep{Abdo09,Ackermann10,Li10}.

\begin{figure}
  \begin{center}
    \includegraphics[width=0.46\textwidth]{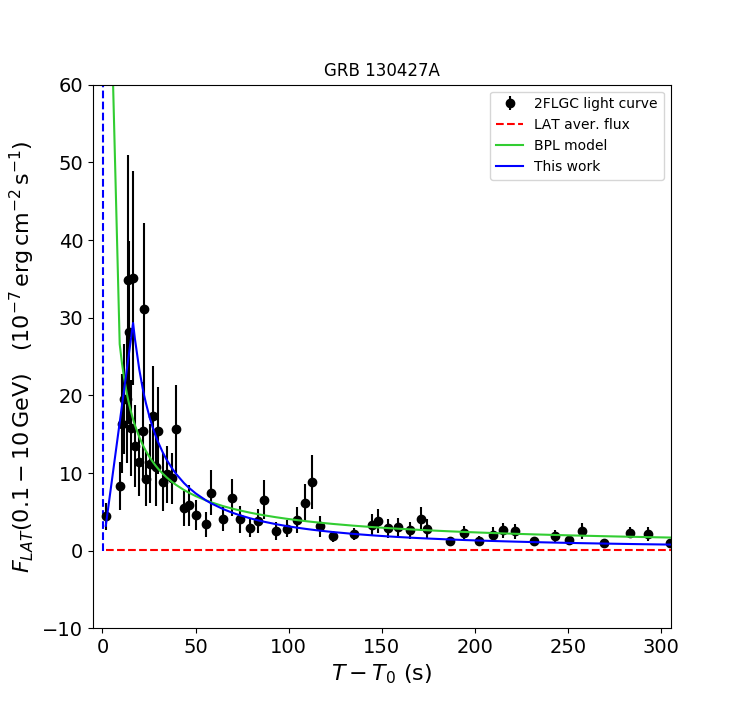}
  \end{center}
  \caption{Energy flux light-curve of GRB~130427A, considering only time intervals with a detection $TS > 9$ (black dots with error bars). The horizontal red dashed line is the average energy flux detected by LAT between $100\,$MeV and $10\,$GeV during the whole burst duration. The vertical blue dashed line represents the start of the LAT analysis. The green continuous line is the broken power-law fit of 2FLGC data, while the blue continuous line is the assumed temporal behaviour in this work. \label{fig01}}
\end{figure}
During the first $10$ years of LAT observations, the {\it Fermi} Gamma-ray Burst Monitor \citep[GBM,][]{GBMpaper} triggered 2357 times, having the source position within the LAT field of view (FoV) for approximately half of the events. This circumstance offers an unprecedented opportunity to study the production of HE radiation and the transition from the prompt emission to the early afterglow phase. While the prompt emission in the lower GBM energy range is generally well interpreted in terms of Band functions \citep{Band93}, the same does not apply to the LAT data. The brightest events, in particular, suggest the existence of an additional component, both spectrally and temporally distinct, arising between $100\,$MeV and $10\,$GeV, which takes the form of a power-law or of a broken power-law function of time, with the occasional detection of photons at $E \geq 10\,$GeV. The time of the onset, the energetics and the evolution of this component are all critical factors whose modelling and interpretation are still under debate. Fig.~\ref{fig01} shows an example of such emission component, for the case of GRB~130427A, a burst detected with a total test statistics $TS = 2794.53$ in the LAT band, and whose emission was tracked with high confidence ($TS > 25$)\footnote{LAT defines the test statistics as twice the logarithmic ratio of the likelihood of a model with a test source at the GRB position with respect to a pure background model, i.e. $TS = 2 \log (\mathcal{L} / \mathcal{L}_0)$. This definition approximately corresponds to the square of detection significance in $\sigma$\ units.} for approximately $1000\,$s after the trigger.

The simple assumption of a power-law behaviour precludes a reasonable fit of the burst at early times, which is a critical point in our study. In order to obtain more realistic estimates of the energy flux radiated by the events, we adopted a modified light-curve representation, where we introduced an increasing flux phase, up to the time of peak emission, followed by a power-law decay. This approach has the advantage of yielding a finite energy flux at early times and it is consistent with theoretical models that require more time to produce HE photons \citep{Asano12}. Since the actual observations of the burst onset are only available for a handful of bright GRBs, while most of the times the peak of the light-curve coincides with the first available point, we assumed a linear flux increase, up to the peak, followed by a best fit power-law decay. The choice of a linear increase is motivated by the scarce amount of information ahead of the peak for most of the observed bursts. This choice may lead to an overestimate of the expected flux in early times, if the peak is much sharper. Due to the typically short duration of this phase, however, the difference between modeled and observed profiles tends to remain low and its effects will be briefly discussed later on. In this way, the energy flux at any instant $t$ is given in the form of:
\begin{equation}
  F_{LAT}(t) = \left\{ \begin{array}{lcl}
    F_p \left( \dfrac{t - T_0}{T_p - T_0} \right) & {\rm for} & T_0 \leq t < T_p \\
     & & \\
    F_p \left( \dfrac{t}{T_p} \right)^{-\alpha} & {\rm for} & t \geq T_p,
  \end{array} \right. \label{eqnLC}
\end{equation}
where $F_{LAT}(t)$ represents the energy flux measured by {\it Fermi}-LAT at instant $t$ between $100\,$MeV and $10\,$GeV, $F_p$ the peak flux, $T_0$ the trigger time, $T_p$ the time needed to achieve maximum flux, expressed in seconds, and $\alpha$\ the power-law index of the time-dependent decay. All the parameters appearing in Eq.~(\ref{eqnLC}) are measured in the 2FLGC catalogue, with the exception of the peak normalization that is computed based on the duration of the LAT signal and the average energy flux.

The choice of the time-dependent flux presented by Eq.~(\ref{eqnLC}) is necessary to investigate the visibility of the events at different times with monitoring or follow-up instruments, but it requires the presence of meaningful light-curve information. While the 2FLGC includes 184 GRBs, not all of them provide the necessary measurements to constrain the light curve. Therefore, we restricted our study to the 172 cases where the LAT analysis obtained time resolved estimates of the photon flux. Since no further cuts were applied and given the large number of GBM triggers that were covered by LAT observations, without a significant detection, we consider that this sample represents an estimate of the fraction of GRBs that can produce HE photons, although the real number of events may be larger, particularly if the energetic emission is commonly concentrated close to the initial stage of the process.

\begin{figure*}
  \begin{center}
    \includegraphics[width=0.9\textwidth]{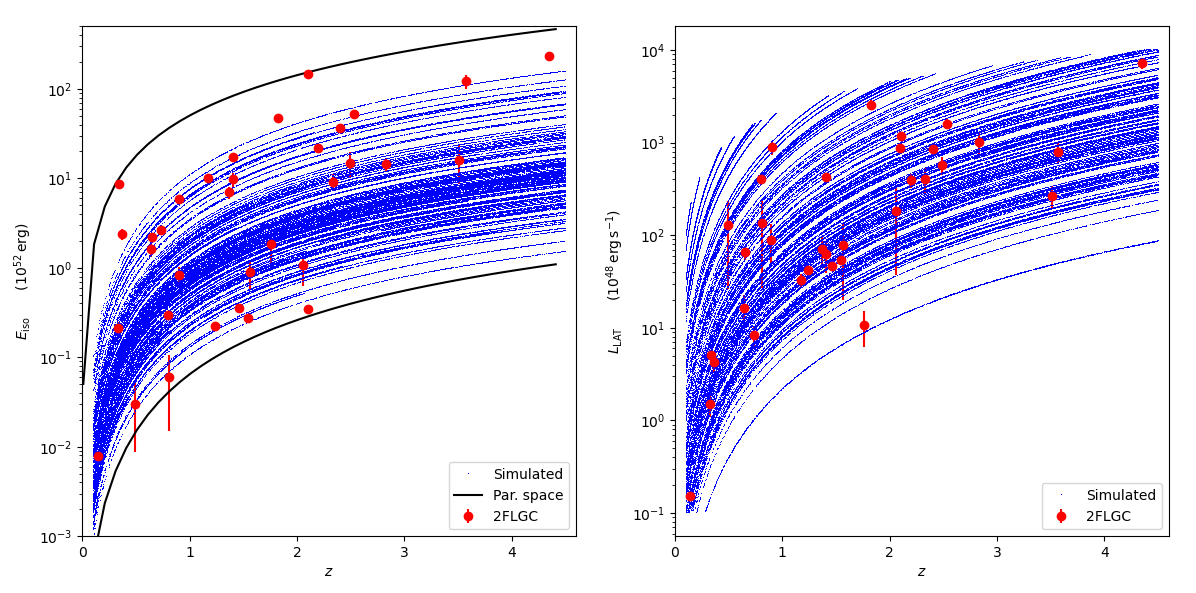}
  \end{center}
  \caption{Comparison of the estimated isotropic energies ($E_{iso}$, left panel) and the luminosities computed in the LAT energy range $0.1\, {\rm GeV} \leq E \leq 10\, {\rm GeV}$ ($L_{LAT}$, right panel) for the sample of LAT detected GRBs with measured redshift (large red points) and the set of simulations carried out on those without a redshift estimate (small blue points). The black curves represent the $E_{iso}$ limits that a GRB with $\gamma = 2.2$ would need to achieve, in order to be observed with the minimum or the maximum LAT fluence at any given redshift. \label{fig02}}
\end{figure*}
\subsection{Extrapolation to the VHE domain}
The spectral and temporal properties of GRBs in the VHE domain are still largely unexplored, having been actually observed only in the afterglow of few events. However, when simultaneous {\it Fermi}-LAT and IACT observations are available, they suggest that the VHE emission exceeds a simple extrapolation of the LAT detected spectrum \citep{GRBpaper,GRBpaper2,Acciari21}. As a consequence, the data included in 2FLGC form a good basis to estimate VHE flux expectations. For our analysis, in particular, we used the reported average photon and energy fluxes, the spectral index of the best power-law fit, the time lag of the LAT signal, the redshift, if available, and the energy flux points of the light-curve. We assume that the differential photon spectra of the GRBs can be described by power-laws with extra-galactic background light (EBL) absorption, according to:
\begin{equation}
  \dfrac{\de N(t)}{\de E} = N_0(t) \left( \dfrac{E}{E_0} \right)^{-\gamma} \exp [-\tau(E, z)] \: \mathrm{ph\, GeV^{-1}\, cm^{-2}\, s^{-1}}, \label{eqnSpec}
\end{equation}
where $N_0(t)$ is the photon flux at scaling energy $E_0$, $\gamma$\ is the photon index for the best power-law fit to the LAT data, and $\tau(E, z)$ is the EBL opacity at energy $E$ and redshift $z$ \citep{Dominguez11}. The normalization of the spectra of GRBs was obtained by forcing the fluxes resulting from the integration of Eq.~(\ref{eqnSpec}) over time and energy between $100\,$MeV and $10\,$GeV to match the values reported in 2FLGC, with the assumption of a light-curve represented by Eq.~(\ref{eqnLC}).

In principle, we could directly use Eq.~(\ref{eqnSpec}) to extrapolate the spectra to the VHE domain, and thus infer the expected VHE GRB integral fluxes. In practice, this operation is not straightforward, because the redshift of the 2FLGC GRBs is actually known only in 34 cases. The majority of these events are observed at redshift $z > 1$ and, though the effects of EBL opacity on the LAT energy band are still very small, their consequences at higher energies are much more severe and they need to be properly accounted for. Since EBL opacity is very likely to suppress most of the VHE signal for $z > 1$ and the number of GRBs with a measured low redshift is too small to apply useful constraints on the rate of events with a detectable HE component, we followed a {\it Monte Carlo} approach to estimate what fraction of the unknown redshift events has realistic chances to occur within the EBL opacity horizon. The problem that we need to address is that the observed flux distribution can result from different combinations of GRB luminosity and redshift, with different effects on the VHE extrapolation. We, therefore, sought for a solution that could express the probability of having a GRB of measured flux at a specific redshift and, therefore, being subject to a certain amount of EBL opacity. The core of our method resides in the assignment of $1000$ random redshift values, distributed on the same $0.1-4.5$ range of measured ones, to each of the $138$ events without a redshift measurement. In order to take into account all the signal collected by LAT, we used the measured fluence to estimate the corresponding intrinsic isotropic energy $E_{iso}$. For a GRB located at redshift $z$, the isotropic energy in the LAT band $E_{iso}$ is given by:
\begin{equation}
  E_{iso} = \dfrac{4 \pi d_L^2}{1 + z} \left[ (T_1 - T_0) \int_{E_1 / (1 + z)}^{E_2 / (1 + z)} E \dfrac{\de N}{\de E} \de E \right], \label{eqnEiso}
\end{equation}
where $d_L$ is the luminosity distance corresponding to $z$, $T_0$ and $T_1$ are the start and the end times of the interval covered by the LAT signal, $E_1 = 0.1\,$GeV and $E_2 = 10\,$GeV are the energy boundaries used in the LAT GRB analysis, and the term in square brackets equals the measured LAT fluence. Eq.~(\ref{eqnEiso}) shows that, for any given fluence, the predicted value of $E_{iso}$ is solely a function of redshift. Recalling that the differential energy spectrum used in Eq.~(\ref{eqnEiso}) is given by Eq.~(\ref{eqnSpec}), the effects of EBL opacity are also roughly taken into account. In the comparison of observed and simulated events, however, the opacity only plays a secondary role, because it remains at a negligible level for the {\it Fermi}-LAT band, except for the highest redshift events. This circumstance has no effect on the following considerations, because any spectral extrapolation to the VHE domain is strongly suppressed already at much lower redshifts.

\begin{figure}
  \begin{center}
    \includegraphics[width=0.48\textwidth]{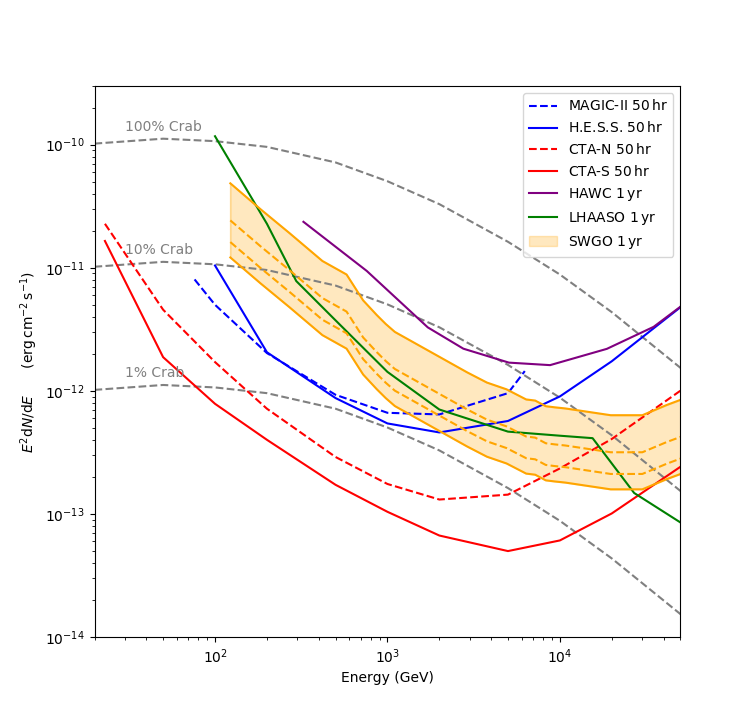}
  \end{center}
  \caption{Differential sensitivity curves to a point-like source in the VHE domain for present day and next-generation IACT observatories. We also plot for comparison the sensitivity of HAWC and LHAASO, together with a grid of sensitivities, used in this work to evaluate the GRB detection potential of a new EAS array, as investigated by the SWGO collaboration \citep{SWGOpaper2}. All sensitivities refer to a steady point-like source, observed at a zenith distance $\theta = 20^{\rm o}$. Different fractions of the Crab Nebula standard flux are plotted as grey dashed lines, for reference. \label{fig03}}
\end{figure}

By repeating our simulations $1000$ times, we obtain a corresponding set of GRBs with their LAT measured spectral and temporal properties, but arranged in different luminosity and redshift distributions, all of which result in the observed fluences. As it is shown in Fig.~\ref{fig03}, the simulated GRBs define a series of tracks that do not uniformly fill the $E_{iso}$ vs. $z$ plane, because each one is constrained by its measured fluence. The two curves that delimit the range of simulated events correspond, respectively, to the $E_{iso}$ of a burst having the lowest fluence measured by LAT, in the assumption of a power-law spectrum with photon index $\gamma = 2.2$ (a value lying between the average index $\gamma_{avg}= 2.22$ and the median index $\gamma_{med} = 2.08$ of 2FLGC), and to the constraint of having:
\begin{equation}
  \dfrac{E_{iso}}{10^{50}\, \mathrm{erg}} \leq 5040\, z^{3/2},
\end{equation}
which is equivalent to fixing a maximum intrinsic energy $E_{iso} \leq 5 \cdot 10^{54}\,$erg within $z \leq 4.5$, while still being consistent with the distribution of the observed values. 

The final product of the simulations is a set of $1000$ redshift distributions of events, from which we can estimate the likelihood that GRBs with proper spectral and temporal requirements occur within a redshift limit that is not yet producing severe absorption on the VHE band. This likelihood can be simply obtained as the average of the ratio between the number of potentially visible GRBs and the total number of observed events, taken from all the simulated distributions. If we compare the luminosity range covered by the simulated GRBs, recalling that all the other properties are kept as measured in 2FLGC, we find that it is in fairly good agreement with the distribution of events with known redshift. In particular, we see from the right panel of Fig.~\ref{fig03} that the simulations do not predict an excess of low luminosity events at very high redshift, which would very likely fall below the {\it Fermi}-LAT detection capabilities, while a minority of possibly high luminosity events can be observed at low redshift, a situation that can be observed, particularly in the case of signals with short duration (below few seconds), which leads to poorly constrained determinations of the average flux.

\section{Observation of VHE transients}
The analysis of {\it Fermi}-LAT data clearly demonstrates that GRBs can produce photons with $E \geq 10\,$GeV. More recently, IACT observations detected VHE photons with $E > 300\,$GeV from as soon as $20\,$s after the trigger, up to some hours in the afterglow of the burst \citep{GRBpaper, GRBpaper2}. The general properties of VHE emission in GRBs, however, are not actually well constrained by these observations. Indeed, while the intrinsically low photon fluxes in the VHE range require large instrumented areas, which cannot be covered by a space-borne detector, the extremely fast evolution of the prompt stage implies that the time taken by instruments with a small FoV to point at the source leads to an unavoidable loss of information, concerning the early spectral properties of the event.
Imaging telescopes, such as the Major Atmospheric Gamma-ray Imaging Cherenkov \citep[MAGIC,][]{MAGICpaper}, the Very Energetic Radiation Imaging Telescope Array System \citep[VERITAS,][]{VERITASpaper}, or the High Energy Stereoscopic System \citep[H.E.S.S.,][]{HESSpaper}, achieve the best sensitivity to compact and point-like sources, due to their fair spatial resolution and their excellent background rejection power. However, the requirement to operate only in clear, dark sky conditions makes the investigation of large sky areas and the execution of regular monitoring activities an extremely hard task. It is expected that the construction of the full CTA observatory will result in a remarkable improvement of the performance of IACT facilities, but the only viable option for the execution of regular monitoring activities of wide sky areas relies on EAS based instruments.

Since EAS arrays can extend over large areas, they are able to collect an appreciable number of photons even from events with relatively low flux. In addition, they can operate almost continuously and they cover a rather large FoV ($\sim 1\,$sr), providing optimal monitoring capabilities. The High Altitude Water Cherenkov array \citep[HAWC,][]{HAWCpaper} and the Large High Altitude Air Shower Observatory \citep[LHAASO,][]{LHAASOpaper} are examples of how EAS detector arrays can map large areas of the sky in a continuous manner, becoming suitable candidates for the detection of transients. Although their lower spatial resolution and weaker background rejection capabilities result in worse point-like sensitivity, if not for the highest energies, the application of optimized detector technologies and the extension to large areas gives them the potential to reach attractive performance in terms of limiting flux, as summarized in Fig.~\ref{fig03}.\footnote{\texttt{https://magic.mpp.mpg.de/uploads/pics/info.txt}}$^,$\footnote{\texttt{https://www.mpi-hd.mpg.de/hfm/HESS/}}$^,$\footnote{\texttt{https://www.cta-observatory.org/science/cta-performance/}}$^,$\footnote{\texttt{https://www.hawc-observatory.org/}}$^,$\footnote{\texttt{http://english.ihep.cas.cn/lhaaso/}} At present, however, the only EAS arrays sensitive to VHE gamma-rays are located in the Northern hemisphere, leaving almost the whole Southern sky not covered by any monitoring facility.

The SWGO Collaboration \citep[][]{SWGOpaper} has been established to design and build a novel wide-field EAS array in the Southern hemisphere, therefore extending the monitored area to almost the whole celestial sphere. The SWGO collaboration seeks to investigate new technologies for the array which, among other goals, aim to render it sensitive to the lowest energies ever achieved by this kind of instrument, opening up new prospects for the study of transients, and in particular GRBs. 
Taking as reference the performance that could be reached by an EAS detector array covering an area $80\,000\, \mathrm{m^2}$ with high fill factor ($\gtrsim 80\%$), high time resolution ($\Delta t \simeq 2\,$ns) and low particle energy threshold for detector activation ($E_{th}\simeq 20\,$MeV) \citep{LATTESpaper}, we investigated the fraction of visible GRBs from a population like the one that LAT detected in 10 years. For this purpose, we used the evidence provided by LAT and VHE simultaneous observations to assume that the temporal and spectral properties, measured by LAT up to $E = 10\,$GeV, could be extrapolated to the TeV energy scale, with the application of the effects of EBL opacity. We compared the flux predicted for our sample of GRBs with the sensitivities based on the instrumental concept outlined above, using a low energy threshold in the range between $125\,$GeV and $500\,$GeV and assuming that, on short time-scales, the instrument response is approximately scaling with the square root of the observing time. Using these definitions for the incoming flux and the instrument sensitivity as a function of time, we computed the number of events that are potentially visible and the time required to detect them. We used the HE light-curves in energy flux units, extrapolated to the VHE domain, to estimate the number of GRBs that are associated with an integrated energy flux larger than the corresponding integrated sensitivity, above the considered energy threshold. In addition, we calculated the time required by different instrumental configurations to trigger on the incoming flux. In order to deal with the expected degradation of sensitivity, for events occurring at less favorable zenith distances than the $20^{\rm o}$ reference sensitivity, we ran our calculations on a grid of possible sensitivities and adopting an increasing value for the low energy threshold, up to $500\,$GeV, thus accounting for the expected loss of signal, away from the central region of the FoV.

\begin{figure*}
  \begin{center}
    \includegraphics[width=0.95\textwidth]{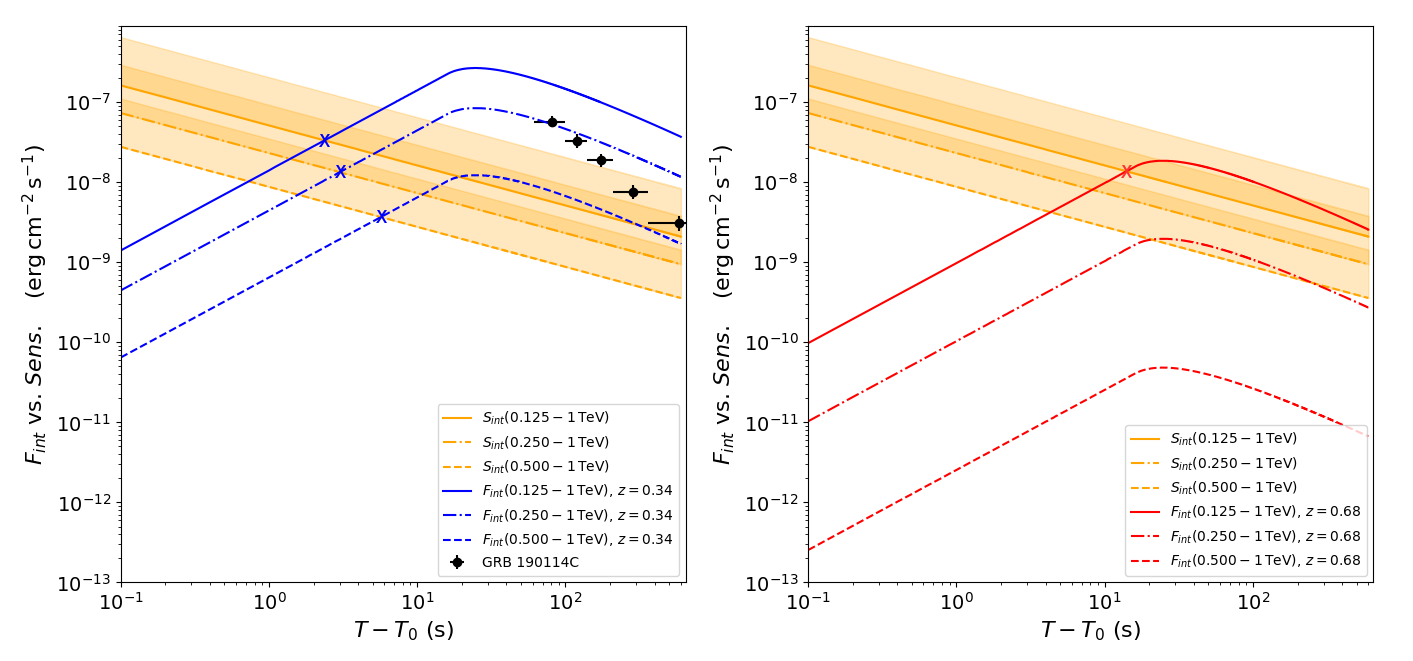}
  \end{center}
  \caption{Integrated energy flux and instrument sensitivity as a function of time for the case of an observed event (GRB~130427A, left panel) and a simulated event (a GRB~130427A-like burst, but located at twice its measured redshift, right panel). The curves represent the average integrated fluxes that would be observed above $125\,$GeV, $250\,$GeV and $500\,$GeV at any given time, while the shaded areas are the corresponding sensitivities. The expected detection times are computed as the times when the average fluxes become larger than the detection limits for the corresponding lower energy thresholds and they are marked as crosses. We also plot, for comparison, the flux measured by MAGIC above $300\,$GeV for the case of GRB~190114C. \label{fig04}}
\end{figure*}
\section{Results and discussion}
\subsection{EAS array GRB detection prospects}
Based on the data collected in this study, we have been able to build an overview of the HE-to-VHE phenomenological timing and spectral properties of the population of LAT-detected GRBs, as inferred from 2FLGC. Our goal is to use such information to predict how frequently the GeV-extrapolated VHE emission of LAT-detected GRBs would be detectable by our model EAS array, and in which observing conditions. Once the spectral and light-curve parameters of the bursts are fixed, and a proper redshift distribution is adopted, we can calculate the predicted VHE fluxes at Earth and compare them with the instrument sensitivities, computed for different exposure times. An example of the process is illustrated in Fig.~\ref{fig04}, where the average integrated fluxes, defined as:
\begin{equation}
  F_{int}(T) = \dfrac{1}{T - T_0} \int_{T_0}^{T} F_{VHE}(t) \de t,
\end{equation}
are compared with a range of integral sensitivities computed above $125\,$GeV, $250\,$GeV and $500\,$GeV. The term $F_{VHE}(t)$ represents the VHE extrapolation above the same energy thresholds of the fluxes expressed by Eq.~(\ref{eqnLC}). In this way, we are able to estimate the number of events, which we expect to be visible with different instrument responses. As it is illustrated in Fig.~\ref{fig04}, the possibility to detect a burst and the required detection time are functions of the burst light-curve, its spectrum and its redshift. The last parameter, in particular, takes a critical role, because integration on higher energies implies a dramatic reduction of incoming flux, due to the EBL term. This, in turn, results in longer detection times for bright, nearby events and it may imply the loss of events originated at higher redshifts, whose fluxes would not attain the corresponding detection limits.

Table~\ref{tabRes} provides a summary of the visibility calculations carried out on our reference GRB population. There we compare the expected number of detections achievable for different instrument configurations, based on $10$ years of data taking. We are careful to distinguish between events with a measured redshift and events following a synthetic redshift distribution. For the simulated redshifts, we use the combination of $1000$ different simulations to estimate the average number of events, at unknown redshift, from which we expect to receive a detectable VHE flux, and the corresponding standard deviations. We subsequently add the real event distribution to the simulated one, to obtain an estimate of the total amount of observable events, with the expected uncertainty. Comparing the assumed instrument sensitivities with the calculated fluxes, both taken as a function of time, we are also able to estimate the average detection times, required to trigger on typical bursts, as well as the corresponding minimum and maximum exposure times needed to detect, respectively, the brightest and the faintest GRB visible to the instrument. This approach and its associated results are likely conservative, due to the excess VHE radiation with respect to a simple extrapolation of the LAT spectrum \citep{GRBpaper,GRBpaper2,Acciari21} and due to the still unconstrained VHE emission properties at early times.

\subsection{Dependence on instrument performance}
To evaluate the detection possibilities of a GRB transient we need to consider the temporal evolution of both the instrumental sensitivity and the source's flux. In our case, in addition to the differential sensitivity, considerations about the detection energy threshold are particularly important. An instrument with a point-source limiting flux of the order of $10^{-11}\, \ergcmsec$ over $1$ year of observing time in the energy range between $100\,$GeV and $500\,$GeV, such as the range of configurations shown in Fig.~\ref{fig03}, can operate with fluxes of order of $10^{-8}\, \ergcmsec$ in time-scales of approximately $1$ hour. A realistic visibility estimate, however, needs to take into account the dependence of the instrument performance on different zenith distances across the FoV. A FoV of $1\,$sr extends to a zenith distance of approximately $32^\mathrm{o}$, where the instrument performance is expected to be lower than the the one calculated at $20^\mathrm{o}$. Beyond this limit, the detection of photons with $E \leq 250\,$GeV is unlikely. We therefore, ran our calculations covering a range of sensitivities and with different energy thresholds. These namely correspond to arrays covering areas of $6.9 \cdot 10^4\, \mathrm{m^2}$, $5.6 \cdot 10^4\, \mathrm{m^2}$ and $4.0 \cdot 10^4\, \mathrm{m^2}$, but they are practically useful to account for observations occurring at larger zenith distances than the reference $20^\mathrm{o}$ calculation.

Comparing the distribution of expected VHE fluxes with the assumed grid of instrument performances, we are able to obtain estimates of the detection possibilities, and of the related trigger times, as well as to investigate the consequences of different limiting parameters on the visibility of a realistic GRB population. The reference design adopted in this work, with optimal sensitivity down to a lower energy threshold of $E_{low} = 125\,$GeV, offers very good performance, being able to detect $17.89$ events, over a reference period of $10$ years, leading to the prediction that nearly two events per year should be in the reach of this type of configuration. However, we need to take into account that the number of potentially visible sources descends from the events observed by LAT. Since the detection of the burst peak emission is likely a discriminating visibility factor, we can adopt a value of approximately $2\,$sr for the LAT instantaneous FoV and assume that the total number of visible events is approximately half the value reported in column 5 of Table~\ref{tabRes}. This implies an expected detection rate of $0.8-1.2$ events per year above $125\,$GeV, for a configuration that works down to low energy, thanks to the fact that the rates expected for configurations that reach down to $50$ percent of the optimal design and with energy thresholds up to $250\,$GeV, are still above $1$ event per year, thus granting for a non negligible contribution from off-zenith bursts.

\begin{table*}
  \caption{Number of GRBs with known and simulated redshifts that are predicted to be brighter than the sensitivity limit of the corresponding instrumental configuration, based on the population of events detected by {\it Fermi}-LAT in $10$ years. The table columns report, respectively, (1) the assumed baseline sensitivity, (2) the low energy threshold, (3) the number of GRBs with measured redshift and a detectable VHE extrapolation, (4) the average number of expected detections, with their standard deviations, from the simulated distribution, (5) the total number of expected detections, (6) the minimum time to trigger on the brightest event, (7) the average detection time, and (8) the maximum time needed to detect the faintest event in the reach of the instrument. \label{tabRes}}
  \begin{tabular}{lccccccc}
    \hline
    \hline
    (1) Instrument & (2) $E_{th}$ & (3) Events with & (4) Simulated & (5) Total & (6) Minimum & (7) Average & (8) Maximum \\
    & & known redshift & events & events & detection time & detection time & detection time \\
    \hline
    SWGO 100\% & $125\,$GeV & $6$ & $11.89 \pm 2.31$ & $17.89 \pm 2.31$ & $0.29\, \usec$ & $46.72\, \usec$ & $2921\, \usec$ \\
    SWGO 100\% & $250\,$GeV & $5$ & $ 8.40 \pm 2.14$ & $13.40 \pm 2.14$ & $0.29\, \usec$ & $41.32\, \usec$ & $3652\, \usec$ \\
    SWGO 100\% & $500\,$GeV & $4$ & $ 5.66 \pm 1.93$ & $ 9.66 \pm 1.93$ & $0.29\, \usec$ & $40.40\, \usec$ & $3286\, \usec$ \\
    SWGO 75\% & $125\,$GeV  & $6$ & $10.45 \pm 2.22$ & $16.45 \pm 2.22$ & $0.29\, \usec$ & $46.60\, \usec$ & $3652\, \usec$ \\
    SWGO 75\% & $250\,$GeV  & $5$ & $ 7.36 \pm 2.05$ & $12.36 \pm 2.05$ & $0.29\, \usec$ & $42.13\, \usec$ & $3652\, \usec$ \\
    SWGO 75\% & $500\,$GeV  & $3$ & $ 5.05 \pm 1.79$ & $ 8.05 \pm 1.79$ & $0.29\, \usec$ & $38.30\, \usec$ & $2921\, \usec$ \\
    SWGO 50\% & $125\,$GeV  & $6$ & $ 8.66 \pm 2.00$ & $14.66 \pm 2.00$ & $0.29\, \usec$ & $44.47\, \usec$ & $3652\, \usec$ \\
    SWGO 50\% & $250\,$GeV  & $5$ & $ 6.13 \pm 1.91$ & $11.13 \pm 1.91$ & $0.29\, \usec$ & $36.62\, \usec$ & $2418\, \usec$ \\
    SWGO 50\% & $500\,$GeV  & $2$ & $ 4.33 \pm 1.63$ & $ 6.33 \pm 1.63$ & $0.29\, \usec$ & $41.04\, \usec$ & $2921\, \usec$ \\
    SWGO 25\% & $125\,$GeV  & $6$ & $ 5.56 \pm 1.73$ & $11.56 \pm 1.73$ & $0.29\, \usec$ & $37.10\, \usec$ & $1814\, \usec$ \\
    SWGO 25\% & $250\,$GeV  & $3$ & $ 4.32 \pm 1.59$ & $ 7.32 \pm 1.59$ & $0.29\, \usec$ & $33.50\, \usec$ & $2821\, \usec$ \\
    SWGO 25\% & $500\,$GeV  & $2$ & $ 3.03 \pm 1.39$ & $ 5.03 \pm 1.39$ & $0.29\, \usec$ & $27.28\, \usec$ & $2418\, \usec$ \\
    \hline
  \end{tabular}
\end{table*}
A further interesting result is the distribution of the expected detection times, defined as the time required by the various instrument configurations taken into account to accumulate an integral flux above the corresponding energy threshold, as shown by the examples illustrated in Fig.~\ref{fig04}. All the tested instrument configurations are able to detect the brightest GRB in the sample in a time-scale below $1\,$s. This burst is identified with GRB~140619B, one of the highest LAT fluence events without a measured redshift and the only one lasting more than $1\,$s with a hard spectral index ($\gamma = 1.88$). Due to its measured flux, this GRB is constrained by the simulations in the redshift range $0.1 \leq z \leq 0.433$, because otherwise it would have unlikely $E_{iso}$ requirements. The estimated detection time of $0.29\,$s corresponds to the observed peak of the LAT light-curve, implying that the assumption of a linear increase for the burst onset has no large effect on our considerations, because any other prescription would at most push the minimum time to achieve detection of a bright GRB exactly at the time of its peak emission. This suggests that SWGO could both act as a fast alert system, to trigger on VHE emission associated with GRBs, and explore the onset of the most energetic component of their spectra. This domain can not be constrained by {\it Fermi}, which is practically limited to $E_{ph} \leq 100\,$GeV on short transients, and it will not be accessible to any present-day or future IACT facility, which can point to the transient only after it has been located by some other monitoring program.

The distribution of average and maximum detection times, on the other hand, shows a non trivial relation with the tested configuration. The average detection times tend to settle below the minute time-scale, showing a steadily decreasing trend for configurations with higher threshold energy or lower overall sensitivity. This apparently contradictory estimate is actually a consequence of the substantial decrease in the expected number of visible events. Configurations with lower reference performance, indeed, can only detect the brightest events of the distribution, therefore being able to trigger in a short time-scale, but only for a smaller number of events. This result is reflected, although with substantial fluctuation, by the estimated maximum detection times, which correspond to the time required to collect a triggering flux from the faintest GRB visible for a specific configuration. These are time-scales of the order of $1\,$hr that tend to increase if the reduction in performance does not significantly affect the population of visible GRBs, but can also locally decrease if the change in performance implies a limitation to the brightest tail of the event distribution.

\subsection{Constraints from early-time detections}
Some of the bursts detected by LAT present an onset of the GeV emission early in the afterglow or even during the prompt phase. This observation can be interpreted in the framework of several models \citep[e.g.][]{Asano09, Ghisellini10, Asano12}. Unfortunately, the data that we possess at present, concerning the onset of the HE signal, are still very limited, to the extent that no standard way to model a full light-curve has been currently identified. The lack of experimental data on the early properties in the HE and VHE domain poses serious limitations to any observational expectations presented here. On the other hand, it represents a unique scientific opportunity for wide-field VHE instruments such as SWGO. Although it is well established that GRBs can be explained by a combination of synchrotron radiation and inverse-Compton scattered photons from a population of ultra-relativistic particles embedded in a strong magnetic field \citep[e.g.][]{Papathanassiou96,Guetta03}, the details on the nature of the emitting particles, the origin of the seed photon field, and the location of the emitting sites, with respect to the central engine, can only be constrained by a full assessment of the temporal evolution of all the spectral components.

In a leptonic scenario, a simple Synchrotron Self-Compton model could account for photons that are $\sim \gamma_e^2$ times more energetic than the synchrotron radiation. Since the typical Lorentz factors of particles accelerated in a GRB jet range from $\gamma_e \simeq 10^3$, in the case of internal shocks, up to $\gamma_e \simeq 10^5$, for forward shocks, and the synchrotron radiation can peak at around $10\,$keV, the expected scattered photons should lie in the range $10\, \mathrm{GeV} \leq E \leq 100\, \mathrm{TeV}$ \citep{Piran04}. Another intriguing possibility is that GRBs may accelerate protons up to the $10^{20}\,$eV energy scale and be responsible for a synchrotron spectrum and a photo-pion signature up to $300\,$GeV \citep{Vietri97}. The occurrence of hadronic interactions, in particular, could be signaled by simultaneous neutrino observations. The efficiency of neutrino production is expected to depend on the GRB temporal evolution \citep{Bustamante17} and, although so far no {\it IceCube} events have been firmly associated with GRB alerts, the absence of a proper sampling of the VHE domain may affect the search for possible connections \citep{Aartsen16}. Finally, above a few hundred GeV, the presence of spectral features, such as breaks, may help constraining the size and the location of the $\gamma$-ray source, through intrinsic opacity arguments. The different signatures of these processes, which are fundamental to understand the mechanisms at the basis of the GRB engine and the jet formation, are expected to arise very early in the event and, therefore, can only be investigated by means of large FoV instruments that have a higher chance of serendipitous detections.

\subsection{Comparison with previous VHE observations}
The results reported in Table~\ref{tabRes} represent an extrapolation to the VHE regime of the distribution of GRBs as seen by {\it Fermi}-LAT in the GeV band. Given the large duration of the LAT monitoring campaign, the statistical properties of this historical sample can be used to predict the frequency of detectable events. However, as already mentioned, extrapolations from the GeV to the VHE range involve considerable uncertainties that need to be taken into account. The first difficulty is the role of EBL extinction, which critically depends on the real redshift distribution of the emitting GRBs. The second is the actual validity of the spectral extrapolations assumed to the VHE domain, without significant intrinsic breaks below, at least, $500\,$GeV. In order to test whether the approach adopted here is acceptable, we attempt a comparison with the available (if limited) VHE observations.

Taking as a reference the case of the MAGIC detection of GRB~190114C \citep{GRBpaper}, for which the integrated flux between $300\,$GeV and $1\,$TeV has been reported to evolve from $F(t_1) \simeq 5 \cdot 10^{-8}\, \ergcmsec$ at $t_1 = 80\,$s after the trigger, down to $F(t_2) \simeq 6 \cdot 10^{-10} \ergcmsec$ after more than $10^3\,$s, we can estimate how frequently events with a comparable flux are predicted in our analysis, therefore providing a reference term between our methodology and the available observations. The observed spectrum of GRB~190114C, obtained in an exposure time of $2392\,$s, is well represented by a simple power-law, with a spectral index $\gamma_{obs} = 5.43 \pm 0.22$, for which we derive an average integrated flux of $\bar{F}_{int} = 1.46 \cdot 10^{-9}\, \ergcmsec$ between $125\,$GeV and $1\,$TeV. If we compare this flux with the predicted VHE flux distributions of the LAT detected GRBs, we find that 12 GRBs (5 real events and 7.04 on average among the simulated-redshift ones) during $10$ years are predicted to have a VHE flux at least as bright as that of GRB~190114C. As a consequence, our calculations suggest that, on average, $1.2$ events per year should be within the reach of currently operating IACT observatories. Considering that Cherenkov telescopes can only operate for approximately $20$ percent of the time, the predicted rate of VHE emitting bursts is consistent with the detection of $2$ events
in approximately $10$ years with both MAGIC and H.E.S.S. On the other hand, the calculation carried out for an array performing at $25$ percent the reference design, above $250\,$GeV predicts that no more than $0.36$ events per year would be visible, with less than $0.2$ within $20^\mathrm{o}$ of zenith distance. With the instrumental characteristics summarized in Fig.~\ref{fig03}, the computed event rate is consistent with the lack of GRB detections by HAWC, up to present.

\section{Conclusions}
The study of VHE radiation from GRBs has fundamental implications on our current understanding of these extremely energetic transient events. In addition to placing invaluable constraints on the nature of the burst engine and on the interaction of its jets with the environment, it is a probe of Cosmology and Fundamental Physics effects that will play a primary role in the era of multi-messenger astronomical observations. However, many critical questions will only be answered with the ability to directly observe the onset of the VHE emission and to characterize its distribution among the GRB population. Although IACTs offer the best sensitivity in this energy range, their narrow FoV hinders the chances that a GRB transient may be observed early enough. In this context, the monitoring capabilities offered by a wide-field, high duty-cycle instrument provides an attractive complementary approach.

Here we used the HE data from GRBs detected during the first $10$ years of {\it Fermi}-LAT observations, to estimate the contributions that survey instruments based on EAS arrays could bring to the study of these events. Using the fact that GRBs have now been firmly detected in the HE and VHE domain, we extracted spectral and timing information from the properties of the LAT detected events. With the aid of ancillary simulations to derive fiducial redshift distributions for the majority of the sources with unknown $z$, we extrapolated the predicted spectra up to the TeV scale, taking into account the effects of EBL opacity. Using the temporal evolution of the LAT signal to model the light-curves, we calculated the detection prospects and the reaction times of different instrumental configurations. We found that a new monitoring facility, based on the proposed SWGO configuration, would allow for the detection of VHE emission for approximately $1$ event per year, providing, in addition, the unprecedented opportunity to study the burst onset emission in this spectral domain. We concluded our investigation showing that, though the results are robust against small changes in instrument performance, reaching down to $75$ percent of the optimal design above $250\,$GeV, a condition that is more likely applying far from the FoV center, solutions that do not offer good low energy performance are very unlikely to provide a relevant contribution in GRB science. We suggest that a combination of Northern and Southern monitoring instruments, such as SWGO and LHAASO, if able to track $\gamma$-ray sources down to few hundreds GeV, would grant a nearly all-sky coverage, significantly increasing the total detection prospects. In addition to working as effective alert systems, they would open a completely new window to explore the production of VHE photons in GRBs from their earliest evolutionary stages, probing a time and spectral domain that cannot be accessed by any other type of astronomical instrumentation, with clear advantages in the localization of VHE transients and their potential associations in multi-messenger studies.

\section*{Data availability}
There are no new data associated with this article.

\section*{Acknowledgements}
The SWGO Collaboration acknowledges the support from the agencies and organizations listed
here: \url{https://www.swgo.org/SWGOWiki/doku.php?id=acknowledgements}. The authors gratefully thank Anthony Brown and Vitor de Souza for suggesting improvements to this study. Special thanks are also offered to the referee, for the careful revision of the manuscript.

This work was partly performed under project PTDC/FIS-PAR/4300/2020, Funda\c{c}\~ao para a Ci\^encia e Tecnologia. RC is grateful for the financial support by OE - Portugal, FCT, I. P., under DL57/2016/cP1330/cT0002.\\ 

UBdA acknowledges the support of a CNPq Productivity Research Grant no. 311997/2019-8 and a Serrapilheira Institute Grant number Serra - 1812-26906. He also acknowledges the receipt of a FAPERJ Young Scientist Fellowship no. E-26/202.818/2019.\\

The {\it Fermi}-LAT Collaboration acknowledges generous ongoing support from a number of agencies and institutes that have supported both the development and the operation of the LAT as well as scientific data analysis. These include the National Aeronautics and Space Administration and the Department of Energy in the United States, the Commissariat \`a l'Energie Atomique and the Centre National de la Recherche Scientifique / Institut National de Physique Nucl\'eaire et de Physique des Particules in France, the Agenzia Spaziale Italiana and the Istituto Nazionale di Fisica Nucleare in Italy, the Ministry of Education, Culture, Sports, Science and Technology (MEXT), High Energy Accelerator Research Organization (KEK) and Japan Aerospace Exploration Agency (JAXA) in Japan, and the K. A. Wallenberg Foundation, the Swedish Research Council and the Swedish National Space Agency in Sweden.

Additional support for science analysis during the operations phase is gratefully acknowledged from the Istituto Nazionale di Astrofisica in Italy and the Centre National d'\'Etudes Spatiales in France. This work performed in part under DOE Contract DEAC02-76SF00515.

\bibliographystyle{mnras}
\bibliography{biblio}

\begin{thebibliography}{}
\makeatletter
\relax
\def\mn@urlcharsother{\let\do\@makeother \do\$\do\&\do\#\do\^\do\_\do\%\do\~}
\def\mn@doi{\begingroup\mn@urlcharsother \@ifnextchar [ {\mn@doi@}
  {\mn@doi@[]}}
\def\mn@doi@[#1]#2{\def\@tempa{#1}\ifx\@tempa\@empty \href
  {http://dx.doi.org/#2} {doi:#2}\else \href {http://dx.doi.org/#2} {#1}\fi
  \endgroup}
\def\mn@eprint#1#2{\mn@eprint@#1:#2::\@nil}
\def\mn@eprint@arXiv#1{\href {http://arxiv.org/abs/#1} {{\tt arXiv:#1}}}
\def\mn@eprint@dblp#1{\href {http://dblp.uni-trier.de/rec/bibtex/#1.xml}
  {dblp:#1}}
\def\mn@eprint@#1:#2:#3:#4\@nil{\def\@tempa {#1}\def\@tempb {#2}\def\@tempc
  {#3}\ifx \@tempc \@empty \let \@tempc \@tempb \let \@tempb \@tempa \fi \ifx
  \@tempb \@empty \def\@tempb {arXiv}\fi \@ifundefined
  {mn@eprint@\@tempb}{\@tempb:\@tempc}{\expandafter \expandafter \csname
  mn@eprint@\@tempb\endcsname \expandafter{\@tempc}}}

\bibitem[\protect\citeauthoryear{{Aartsen} et~al.,}{{Aartsen}
  et~al.}{2016}]{Aartsen16}
{Aartsen} M.~G.,  et~al., 2016, \mn@doi [\apj] {10.3847/0004-637X/824/2/115},
  \href {https://ui.adsabs.harvard.edu/abs/2016ApJ...824..115A} {824, 115}

\bibitem[\protect\citeauthoryear{{Abbott} et~al.,}{{Abbott}
  et~al.}{2017a}]{Abbott17a}
{Abbott} B.~P.,  et~al., 2017a, \mn@doi [\apjl] {10.3847/2041-8213/aa91c9},
  \href {https://ui.adsabs.harvard.edu/abs/2017ApJ...848L..12A} {848, L12}

\bibitem[\protect\citeauthoryear{{Abbott} et~al.,}{{Abbott}
  et~al.}{2017b}]{Abbott17b}
{Abbott} B.~P.,  et~al., 2017b, \mn@doi [\apjl] {10.3847/2041-8213/aa920c},
  \href {https://ui.adsabs.harvard.edu/abs/2017ApJ...848L..13A} {848, L13}

\bibitem[\protect\citeauthoryear{{Abdalla} \& {et al.}}{{Abdalla} \& {et
  al.}}{2019}]{GRBpaper2}
{Abdalla} H.,  {et al.} 2019, \mn@doi [\nat] {10.1038/s41586-019-1743-9}, \href
  {https://ui.adsabs.harvard.edu/abs/2019Natur.575..464A} {575, 464}

\bibitem[\protect\citeauthoryear{{Abdo} et~al.,}{{Abdo} et~al.}{2009}]{Abdo09}
{Abdo} A.~A.,  et~al., 2009, \mn@doi [Science] {10.1126/science.1169101}, \href
  {https://ui.adsabs.harvard.edu/abs/2009Sci...323.1688A} {323, 1688}

\bibitem[\protect\citeauthoryear{{Acciari} et~al.,}{{Acciari}
  et~al.}{2021}]{Acciari21}
{Acciari} V.~A.,  et~al., 2021, \mn@doi [\apj] {10.3847/1538-4357/abd249},
  \href {https://ui.adsabs.harvard.edu/abs/2021ApJ...908...90A} {908, 90}

\bibitem[\protect\citeauthoryear{{Ackermann} et~al.,}{{Ackermann}
  et~al.}{2010}]{Ackermann10}
{Ackermann} M.,  et~al., 2010, \mn@doi [\apj] {10.1088/0004-637X/716/2/1178},
  \href {https://ui.adsabs.harvard.edu/abs/2010ApJ...716.1178A} {716, 1178}

\bibitem[\protect\citeauthoryear{{Aharonian} et~al.,}{{Aharonian}
  et~al.}{2006}]{HESSpaper}
{Aharonian} F.,  et~al., 2006, \mn@doi [\aap] {10.1051/0004-6361:20065351},
  \href {https://ui.adsabs.harvard.edu/abs/2006A&A...457..899A} {457, 899}

\bibitem[\protect\citeauthoryear{{Ajello} et~al.,}{{Ajello}
  et~al.}{2019}]{2FLGCpaper}
{Ajello} M.,  et~al., 2019, \mn@doi [\apj] {10.3847/1538-4357/ab1d4e}, \href
  {https://ui.adsabs.harvard.edu/abs/2019ApJ...878...52A} {878, 52}

\bibitem[\protect\citeauthoryear{{Aleksi{\'c}} et~al.,}{{Aleksi{\'c}}
  et~al.}{2016}]{MAGICpaper}
{Aleksi{\'c}} J.,  et~al., 2016, \mn@doi [Astroparticle Physics]
  {10.1016/j.astropartphys.2015.04.004}, \href
  {https://ui.adsabs.harvard.edu/abs/2016APh....72...61A} {72, 61}

\bibitem[\protect\citeauthoryear{{Asano} \& {M{\'e}sz{\'a}ros}}{{Asano} \&
  {M{\'e}sz{\'a}ros}}{2012}]{Asano12}
{Asano} K.,  {M{\'e}sz{\'a}ros} P.,  2012, \mn@doi [\apj]
  {10.1088/0004-637X/757/2/115}, \href
  {https://ui.adsabs.harvard.edu/abs/2012ApJ...757..115A} {757, 115}

\bibitem[\protect\citeauthoryear{{Asano}, {Guiriec}  \&
  {M{\'e}sz{\'a}ros}}{{Asano} et~al.}{2009}]{Asano09}
{Asano} K.,  {Guiriec} S.,   {M{\'e}sz{\'a}ros} P.,  2009, \mn@doi [\apjl]
  {10.1088/0004-637X/705/2/L191}, \href
  {https://ui.adsabs.harvard.edu/abs/2009ApJ...705L.191A} {705, L191}

\bibitem[\protect\citeauthoryear{{Assis} et~al.,}{{Assis}
  et~al.}{2018}]{LATTESpaper}
{Assis} P.,  et~al., 2018, \mn@doi [Astroparticle Physics]
  {10.1016/j.astropartphys.2018.02.004}, \href
  {https://ui.adsabs.harvard.edu/abs/2018APh....99...34A} {99, 34}

\bibitem[\protect\citeauthoryear{{Atwood} et~al.,}{{Atwood}
  et~al.}{2009}]{LATpaper}
{Atwood} W.~B.,  et~al., 2009, \mn@doi [\apj] {10.1088/0004-637X/697/2/1071},
  \href {https://ui.adsabs.harvard.edu/abs/2009ApJ...697.1071A} {697, 1071}

\bibitem[\protect\citeauthoryear{{Band} et~al.,}{{Band} et~al.}{1993}]{Band93}
{Band} D.,  et~al., 1993, \mn@doi [\apj] {10.1086/172995}, \href
  {https://ui.adsabs.harvard.edu/abs/1993ApJ...413..281B} {413, 281}

\bibitem[\protect\citeauthoryear{{Barres de Almeida}, {Giacinti}  \&
  {Longo}}{{Barres de Almeida} et~al.}{2021}]{SWGOpaper}
{Barres de Almeida} U.,  {Giacinti} G.,   {Longo} F.,  2021, in 37th
  International Cosmic Ray Conference (ICRC2021). p.~10

\bibitem[\protect\citeauthoryear{{Bennett}, {Larson}, {Weiland}  \&
  {Hinshaw}}{{Bennett} et~al.}{2014}]{Bennett14}
{Bennett} C.~L.,  {Larson} D.,  {Weiland} J.~L.,   {Hinshaw} G.,  2014, \mn@doi
  [\apj] {10.1088/0004-637X/794/2/135}, \href
  {https://ui.adsabs.harvard.edu/abs/2014ApJ...794..135B} {794, 135}

\bibitem[\protect\citeauthoryear{{Blanchard} et~al.,}{{Blanchard}
  et~al.}{2017}]{Blanchard17}
{Blanchard} P.~K.,  et~al., 2017, \mn@doi [\apjl] {10.3847/2041-8213/aa9055},
  \href {https://ui.adsabs.harvard.edu/abs/2017ApJ...848L..22B} {848, L22}

\bibitem[\protect\citeauthoryear{{Bloom} et~al.,}{{Bloom}
  et~al.}{1999}]{Bloom99}
{Bloom} J.~S.,  et~al., 1999, \mn@doi [\nat] {10.1038/46744}, \href
  {https://ui.adsabs.harvard.edu/abs/1999Natur.401..453B} {401, 453}

\bibitem[\protect\citeauthoryear{{Bustamante}, {Heinze}, {Murase}  \&
  {Winter}}{{Bustamante} et~al.}{2017}]{Bustamante17}
{Bustamante} M.,  {Heinze} J.,  {Murase} K.,   {Winter} W.,  2017, \mn@doi
  [\apj] {10.3847/1538-4357/837/1/33}, \href
  {https://ui.adsabs.harvard.edu/abs/2017ApJ...837...33B} {837, 33}

\bibitem[\protect\citeauthoryear{{CTA Consortium}}{{CTA
  Consortium}}{2019}]{CTApaper}
{CTA Consortium} 2019, {Science with the Cherenkov Telescope Array},
  \mn@doi{10.1142/10986.
}

\bibitem[\protect\citeauthoryear{{Cano}}{{Cano}}{2016}]{Cano16}
{Cano} Z.,  2016, in Eighth Huntsville Gamma-Ray Burst Symposium. p.~4116

\bibitem[\protect\citeauthoryear{{Cao} et~al.,}{{Cao}
  et~al.}{2019}]{LHAASOpaper}
{Cao} Z.,  et~al., 2019, \mn@doi [\caa] {10.1016/j.chinastron.2019.11.001},
  \href {https://ui.adsabs.harvard.edu/abs/2019ChA&A..43..457C} {43, 457}

\bibitem[\protect\citeauthoryear{{DeYoung}}{{DeYoung}}{2012}]{HAWCpaper}
{DeYoung} T.,  2012, \mn@doi [Nuclear Instruments and Methods in Physics
  Research A] {10.1016/j.nima.2012.01.026}, \href
  {https://ui.adsabs.harvard.edu/abs/2012NIMPA.692...72D} {692, 72}

\bibitem[\protect\citeauthoryear{{Dom{\'\i}nguez} et~al.,}{{Dom{\'\i}nguez}
  et~al.}{2011}]{Dominguez11}
{Dom{\'\i}nguez} A.,  et~al., 2011, \mn@doi [\mnras]
  {10.1111/j.1365-2966.2010.17631.x}, \href
  {https://ui.adsabs.harvard.edu/abs/2011MNRAS.410.2556D} {410, 2556}

\bibitem[\protect\citeauthoryear{{Eichler}, {Livio}, {Piran}  \&
  {Schramm}}{{Eichler} et~al.}{1989}]{Eichler89}
{Eichler} D.,  {Livio} M.,  {Piran} T.,   {Schramm} D.~N.,  1989, \mn@doi
  [\nat] {10.1038/340126a0}, \href
  {https://ui.adsabs.harvard.edu/abs/1989Natur.340..126E} {340, 126}

\bibitem[\protect\citeauthoryear{{Fishman} \& {Meegan}}{{Fishman} \&
  {Meegan}}{1995}]{Fishman95}
{Fishman} G.~J.,  {Meegan} C.~A.,  1995, \mn@doi [\araa]
  {10.1146/annurev.aa.33.090195.002215}, \href
  {https://ui.adsabs.harvard.edu/abs/1995ARA&A..33..415F} {33, 415}

\bibitem[\protect\citeauthoryear{{Fruchter} et~al.,}{{Fruchter}
  et~al.}{2006}]{Fruchter06}
{Fruchter} A.~S.,  et~al., 2006, \mn@doi [\nat] {10.1038/nature04787}, \href
  {https://ui.adsabs.harvard.edu/abs/2006Natur.441..463F} {441, 463}

\bibitem[\protect\citeauthoryear{{Galama} et~al.,}{{Galama}
  et~al.}{1998}]{Galama98}
{Galama} T.~J.,  et~al., 1998, \mn@doi [\nat] {10.1038/27150}, \href
  {https://ui.adsabs.harvard.edu/abs/1998Natur.395..670G} {395, 670}

\bibitem[\protect\citeauthoryear{{Ghisellini}, {Ghirlanda}, {Nava}  \&
  {Celotti}}{{Ghisellini} et~al.}{2010}]{Ghisellini10}
{Ghisellini} G.,  {Ghirlanda} G.,  {Nava} L.,   {Celotti} A.,  2010, \mn@doi
  [\mnras] {10.1111/j.1365-2966.2009.16171.x}, \href
  {https://ui.adsabs.harvard.edu/abs/2010MNRAS.403..926G} {403, 926}

\bibitem[\protect\citeauthoryear{{Gompertz}, {Fruchter}  \& {Pe'er}}{{Gompertz}
  et~al.}{2018}]{Gompertz18}
{Gompertz} B.~P.,  {Fruchter} A.~S.,   {Pe'er} A.,  2018, \mn@doi [\apj]
  {10.3847/1538-4357/aadba8}, \href
  {https://ui.adsabs.harvard.edu/abs/2018ApJ...866..162G} {866, 162}

\bibitem[\protect\citeauthoryear{{Guetta} \& {Granot}}{{Guetta} \&
  {Granot}}{2003}]{Guetta03}
{Guetta} D.,  {Granot} J.,  2003, \mn@doi [\apj] {10.1086/346221}, \href
  {https://ui.adsabs.harvard.edu/abs/2003ApJ...585..885G} {585, 885}

\bibitem[\protect\citeauthoryear{{Klebesadel}, {Strong}  \&
  {Olson}}{{Klebesadel} et~al.}{1973}]{Klebesadel73}
{Klebesadel} R.~W.,  {Strong} I.~B.,   {Olson} R.~A.,  1973, \mn@doi [\apjl]
  {10.1086/181225}, \href
  {https://ui.adsabs.harvard.edu/abs/1973ApJ...182L..85K} {182, L85}

\bibitem[\protect\citeauthoryear{{Kouveliotou}, {Meegan}, {Fishman}, {Bhat},
  {Briggs}, {Koshut}, {Paciesas}  \& {Pendleton}}{{Kouveliotou}
  et~al.}{1993}]{Kouveliotou93}
{Kouveliotou} C.,  {Meegan} C.~A.,  {Fishman} G.~J.,  {Bhat} N.~P.,  {Briggs}
  M.~S.,  {Koshut} T.~M.,  {Paciesas} W.~S.,   {Pendleton} G.~N.,  1993,
  \mn@doi [\apjl] {10.1086/186969}, \href
  {https://ui.adsabs.harvard.edu/abs/1993ApJ...413L.101K} {413, L101}

\bibitem[\protect\citeauthoryear{{Kulkarni} et~al.,}{{Kulkarni}
  et~al.}{1998}]{Kulkarni98}
{Kulkarni} S.~R.,  et~al., 1998, \mn@doi [\nat] {10.1038/29927}, \href
  {https://ui.adsabs.harvard.edu/abs/1998Natur.393...35K} {393, 35}

\bibitem[\protect\citeauthoryear{{Li}}{{Li}}{2010}]{Li10}
{Li} Z.,  2010, \mn@doi [\apj] {10.1088/0004-637X/709/1/525}, \href
  {https://ui.adsabs.harvard.edu/abs/2010ApJ...709..525L} {709, 525}

\bibitem[\protect\citeauthoryear{{Li} \& {Paczy{\'n}ski}}{{Li} \&
  {Paczy{\'n}ski}}{1998}]{Li98}
{Li} L.-X.,  {Paczy{\'n}ski} B.,  1998, \mn@doi [\apjl] {10.1086/311680}, \href
  {https://ui.adsabs.harvard.edu/abs/1998ApJ...507L..59L} {507, L59}

\bibitem[\protect\citeauthoryear{{MAGIC Collaboration}}{{MAGIC
  Collaboration}}{2019}]{GRBpaper}
{MAGIC Collaboration} 2019, \mn@doi [\nat] {10.1038/s41586-019-1750-x}, \href
  {https://ui.adsabs.harvard.edu/abs/2019Natur.575..455M} {575, 455}

\bibitem[\protect\citeauthoryear{{Mazets} et~al.,}{{Mazets}
  et~al.}{1981}]{Mazets81}
{Mazets} E.~P.,  et~al., 1981, \mn@doi [\apss] {10.1007/BF00649140}, \href
  {https://ui.adsabs.harvard.edu/abs/1981Ap&SS..80....3M} {80, 3}

\bibitem[\protect\citeauthoryear{{Meegan} et~al.,}{{Meegan}
  et~al.}{2009}]{GBMpaper}
{Meegan} C.,  et~al., 2009, \mn@doi [\apj] {10.1088/0004-637X/702/1/791}, \href
  {https://ui.adsabs.harvard.edu/abs/2009ApJ...702..791M} {702, 791}

\bibitem[\protect\citeauthoryear{{Metzger}, {Djorgovski}, {Kulkarni},
  {Steidel}, {Adelberger}, {Frail}, {Costa}  \& {Frontera}}{{Metzger}
  et~al.}{1997}]{Metzger97}
{Metzger} M.~R.,  {Djorgovski} S.~G.,  {Kulkarni} S.~R.,  {Steidel} C.~C.,
  {Adelberger} K.~L.,  {Frail} D.~A.,  {Costa} E.,   {Frontera} F.,  1997,
  \mn@doi [\nat] {10.1038/43132}, \href
  {https://ui.adsabs.harvard.edu/abs/1997Natur.387..878M} {387, 878}

\bibitem[\protect\citeauthoryear{{Nakar} \& {Piran}}{{Nakar} \&
  {Piran}}{2002}]{Nakar02}
{Nakar} E.,  {Piran} T.,  2002, \mn@doi [\mnras]
  {10.1046/j.1365-8711.2002.05136.x}, \href
  {https://ui.adsabs.harvard.edu/abs/2002MNRAS.330..920N} {330, 920}

\bibitem[\protect\citeauthoryear{{Norris}, {Cline}, {Desai}  \&
  {Teegarden}}{{Norris} et~al.}{1984}]{Norris84}
{Norris} J.~P.,  {Cline} T.~L.,  {Desai} U.~D.,   {Teegarden} B.~J.,  1984,
  \mn@doi [\nat] {10.1038/308434a0}, \href
  {https://ui.adsabs.harvard.edu/abs/1984Natur.308..434N} {308, 434}

\bibitem[\protect\citeauthoryear{{Papathanassiou} \&
  {Meszaros}}{{Papathanassiou} \& {Meszaros}}{1996}]{Papathanassiou96}
{Papathanassiou} H.,  {Meszaros} P.,  1996, \mn@doi [\apjl] {10.1086/310343},
  \href {https://ui.adsabs.harvard.edu/abs/1996ApJ...471L..91P} {471, L91}

\bibitem[\protect\citeauthoryear{{Piran}}{{Piran}}{2004}]{Piran04}
{Piran} T.,  2004, \mn@doi [Reviews of Modern Physics]
  {10.1103/RevModPhys.76.1143}, \href
  {https://ui.adsabs.harvard.edu/abs/2004RvMP...76.1143P} {76, 1143}

\bibitem[\protect\citeauthoryear{{Planck Collaboration}}{{Planck
  Collaboration}}{2016}]{PlanckColl16}
{Planck Collaboration} 2016, \mn@doi [\aap] {10.1051/0004-6361/201525830},
  \href {https://ui.adsabs.harvard.edu/abs/2016A&A...594A..13P} {594, A13}

\bibitem[\protect\citeauthoryear{{Schoorlemmer}, {Concei\c{c}\~ao}  \&
  {Smith}}{{Schoorlemmer} et~al.}{2021}]{SWGOpaper2}
{Schoorlemmer} H.,  {Concei\c{c}\~ao} R.,   {Smith} A.~J.,  2021, in 37th
  International Cosmic Ray Conference (ICRC2021). p.~9

\bibitem[\protect\citeauthoryear{{Vietri}}{{Vietri}}{1997}]{Vietri97}
{Vietri} M.,  1997, \mn@doi [\prl] {10.1103/PhysRevLett.78.4328}, \href
  {https://ui.adsabs.harvard.edu/abs/1997PhRvL..78.4328V} {78, 4328}

\bibitem[\protect\citeauthoryear{{Weekes} et~al.,}{{Weekes}
  et~al.}{2002}]{VERITASpaper}
{Weekes} T.~C.,  et~al., 2002, \mn@doi [Astroparticle Physics]
  {10.1016/S0927-6505(01)00152-9}, \href
  {https://ui.adsabs.harvard.edu/abs/2002APh....17..221W} {17, 221}

\bibitem[\protect\citeauthoryear{{Woosley}}{{Woosley}}{1993}]{Woosley93}
{Woosley} S.~E.,  1993, \mn@doi [\apj] {10.1086/172359}, \href
  {https://ui.adsabs.harvard.edu/abs/1993ApJ...405..273W} {405, 273}

\makeatother
\end{thebibliography}
\end{document}